\documentclass[twocolumn]{emulateapj}
\usepackage{epsf}
\usepackage{amsmath}
\usepackage{multirow}
\usepackage{rotating}
\usepackage{subfigure}
\usepackage{multirow}
\usepackage{natbib}
\usepackage{color}
\def\approxprop{%
  \def\p{%
    \setbox0=\vbox{\hbox{$\propto$}}%
    \ht0=0.6ex \box0 }%
  \def\s{%
    \vbox{\hbox{$\sim$}}%
  }%
  \mathrel{\raisebox{0.7ex}{%
      \mbox{$\underset{\s}{\p}$}%
    }}%
}
\begin{document}

\slugcomment{Accepted to ApJ Letters on 18 July 2013}
\shortauthors{A. L. King et al.}
\shorttitle{Disk-Jet Connection in NGC 4395}

\title{A Distinctive Disk-Jet Coupling in the Lowest-Mass Seyfert, NGC 4395}

\author{Ashley~L.~King\altaffilmark{1},
         Jon~M.~Miller\altaffilmark{1},
	Mark~T.~Reynolds\altaffilmark{1},
	Kayhan~G\"ultekin\altaffilmark{1},
	Elena~Gallo\altaffilmark{1},
	Dipankar~Maitra\altaffilmark{1} }
        
\altaffiltext{1}{Department of Astronomy, University of Michigan, 500
Church Street, Ann Arbor, MI 48109-1042, ashking@umich.edu}

\label{firstpage}
\begin{abstract}
Simultaneous observations of  X-rays and radio luminosities have been well studied in accreting stellar-mass black holes. These observations are performed in order to understand how mass accretion rates and jetted outflows are linked in these individual systems. Such contemporaneous studies in supermassive black holes (SMBH) are harder to perform, as viscous times scale linearly with mass. However, as NGC 4395 is the lowest known mass Seyfert galaxy, we have used it to examine the simultaneous X-ray ({\it Swift}) and radio (VLA) correlation in a SMBH in a reasonably timed observing campaign. We find that the intrinsic X-ray variability is stronger than the radio variability,  and that the fluxes are only weakly or tentatively coupled, similar to prior results obtained in NGC 4051. If the corona and the base of the jet are one and the same, this may suggest that the corona in radio-quiet AGN filters disk variations, only transferring the strongest and/or most sustained variations into the jet. Further, when both NGC 4395 and NGC 4051 are placed on the stellar-mass $L_X$-$L_R$ plane, they appear to reside on the steeper $L_X$-$L_R$ track. This suggests that SMBHs also follow two distinct tracks just as stellar-mass black holes do, and supports the idea that the same physical disk-jet mechanisms are at play across the mass scale.
\end{abstract}
 
\section{Introduction}
Observations have revealed a plane connecting the X-ray  and radio luminosity  of black holes that spans the mass scale: the fundamental plane of black hole activity \citep[][]{Merloni03,Falcke04,Gultekin09,Plotkin12}. This plane suggests there is an underlying physical mechanism driving the relation, which acts across the mass scale \citep[A similar relation across the mass scale is also seen in black hole disk-winds and is consistent with disk-jet power relations, ][]{King13a}. In the fundamental plane of black hole activity, the black hole mass is thought to set a limit to the amount of power that can be extracted from the system. The  radio luminosity is taken to be a rough proxy for the jet power as the emission is thought to be synchrotron emission along the jet. Finally, the X-ray luminosity is either directly associated with accretion rate \citep[e.g.,][]{Merloni03} or with the base of the jet \citep[e.g.,][]{Falcke04,Plotkin12}. 

In stellar-mass black holes one can examine how individual sources move across this fundamental plane, as the timescales for variations are particularly short \citep[e.g.,][]{Corbel03,Gallo03,Gallo12,Corbel13}. However, the viscous times on which stellar-mass black holes have been probed scale with mass, resulting in relatively long observing campaigns for all but the smallest of SMBHs. 

Fortunately, there are a select few low-mass SMBHs with short enough viscous timescales for a simultaneous observing campaign. In particular, the Seyfert NGC 4395 is the smallest mass SMBH whose mass has been measured with reliable, reverberation mapping techniques. NGC 4395 has a mass of $M_{BH} = (3.6\pm1.1) \times10^{5} M_\odot$ \citep{Peterson05} and is accreting at $<$0.1\% of its Eddington luminosity \citep[e.g.,][]{Shih03}. It has a very variable X-ray flux \citep[e.g.][]{Nardini11} and harbors a compact, non-thermal radio source \citep{Wrobel06}, making it ideal for a simultaneous campaign aimed to examine the disk-jet connection in an SMBH.

We present the results of a nearly-simultaneous {\it Swift} X-ray and Karl G.\ Jansky Very Large Array (VLA) radio observing campaign. We begin with a brief description of the observations taken, followed by a discussion of the correlation between the two bands. We end by discussing the relation of NGC 4395 with other Seyferts as well as its stellar-mass counter parts.

\section{ Observations}
\subsection{Radio}
\begin{figure*}[t]
\centering
\vspace{5cm}
\subfigure[\label{fig:rlc}]{\includegraphics[scale=.9,clip=true,trim= 2cm 5cm 3cm 18cm]{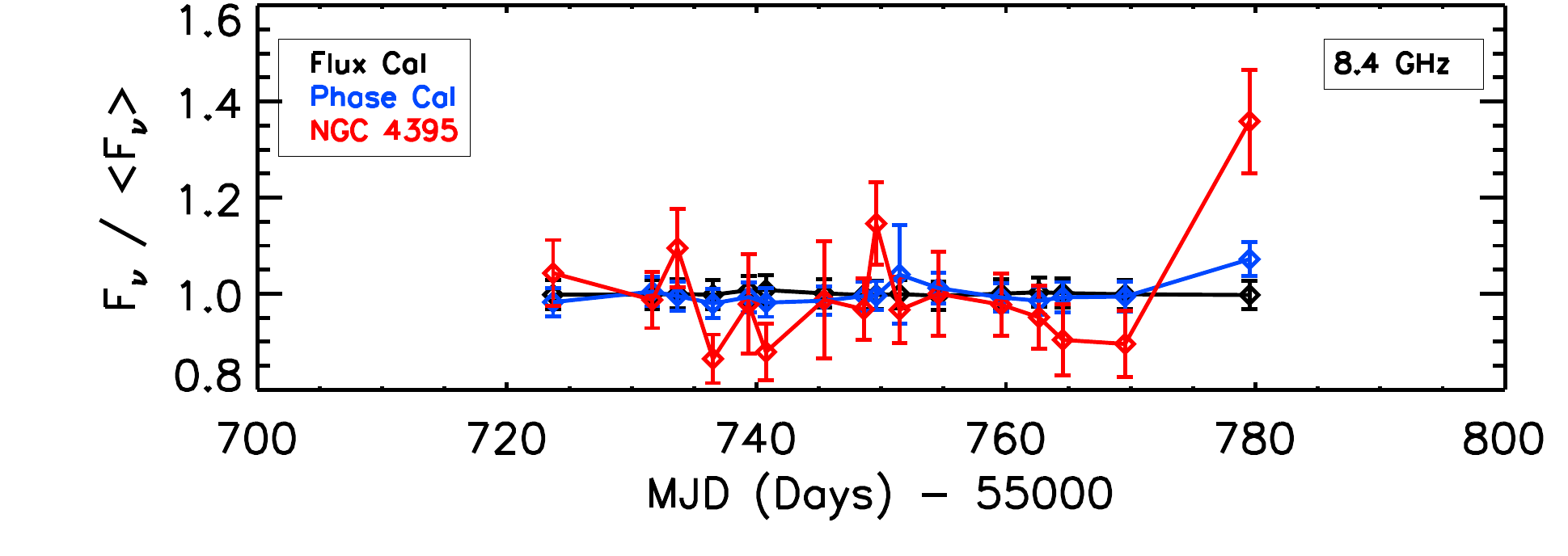}}
\vspace{0cm, \subfigure[\label{fig:lc}]{\includegraphics[scale=.9,angle=0,clip=true,trim= 1cm 5cm 3cm 18cm]{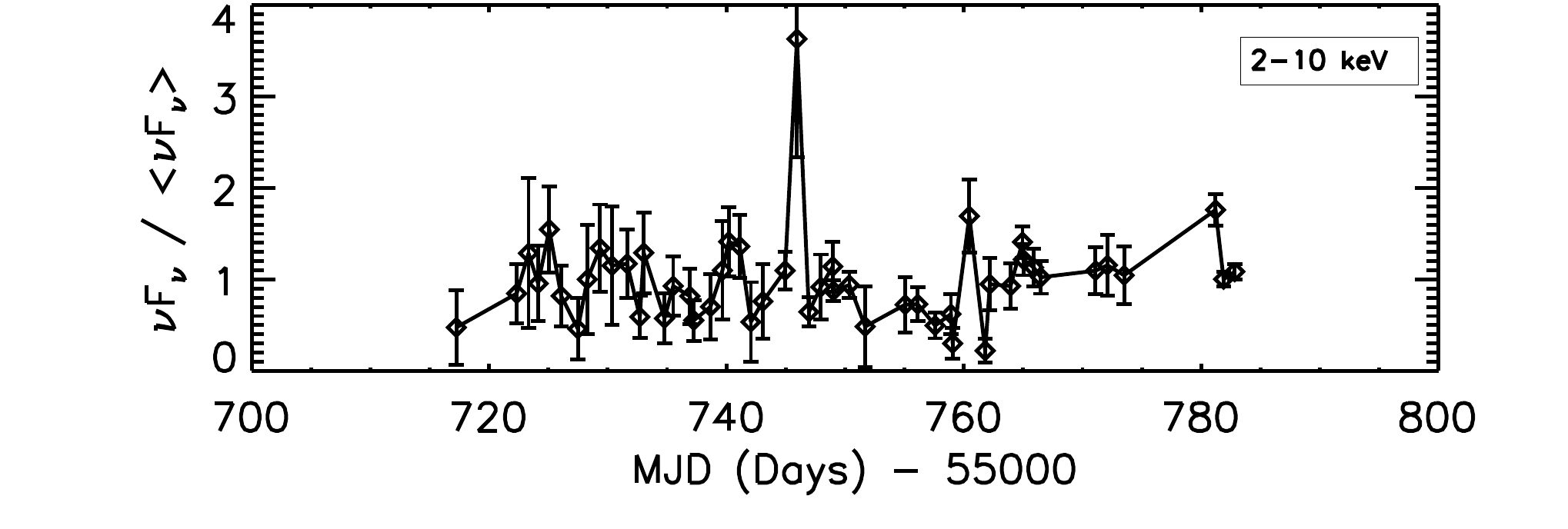} }}
\caption{a) This plot shows the radio variability of NGC 4395 (red) at 8.4 GHz. The flux calibrator (3C 286) is shown in black, and the phase calibrator (J1242+3729) is shown in blue. The flux densities are divided by the mean ($\langle F_R\rangle =5.6\times10^{-4}$ Jy) for easy comparison with the X-ray variability shown in Figure \ref{fig:lc}. b) This plot shows the X-ray variability light curve for the unabsorbed 2--10 keV {\it Swift} band. The X-ray flux is also divided by the mean of the observations, $\langle F_X\rangle =8.2\times10^{-12}$ ergs cm$^{-2}$ s$^{-1}$. }

\end{figure*}

As part of our radio-X-ray monitoring campaign, 16 radio observations were taken with the VLA from 11 June 2011 to 06 August 2011 with an average spacing of 3.7 days. The data were taken at 8.4 GHz with 256 MHz bandwidth in the A configuration. This gave a beam size of $\sim$0.29\arcsec $\times$ 0.26\arcsec. 3C 286 was used as the flux calibrator and J1242+3720 was used as the phase calibrator. We adopted the flux scale Perley-Butler 2010.  Approximately 15 minutes were spent on source, NGC 4395, during each observation. The residuals reached an rms of approximately $2\times10^{-5}$Jy/beam. We used CASA version 3.4.0 \citep{McMullin07} to perform standard flagging, and to create a primary beam corrected image. We utilized the {\it clean} routine with natural weighting of the visibilities. The images showed unresolved point sources and were fit with {\tt imfit}. The errors reported include the observational errors and a 3$\%$ systematic error added in quadrature. Although, this systematic error is relatively small, we note that the phase calibrator in Figure \ref{fig:rlc} shows a vary stable flux density, with an rms at $<3\% \langle F_\nu\rangle$. In addition, the flux calibrator 3C 286 is known to be extremely stable \citep{Perley13}, which is a necessity for this study. The flux-densities are shown in Figure \ref{fig:rlc}, divided by the mean flux-density, which for NGC 4395 is $\langle F_R\rangle =5.6\times10^{-4}$ Jy.

\subsection{X-rays}
\begin{figure*}[t]
\centering
\subfigure[ \label{fig:ctrate}]{\includegraphics[scale=.44,clip=true,trim= 1cm 5cm 3cm 2cm]{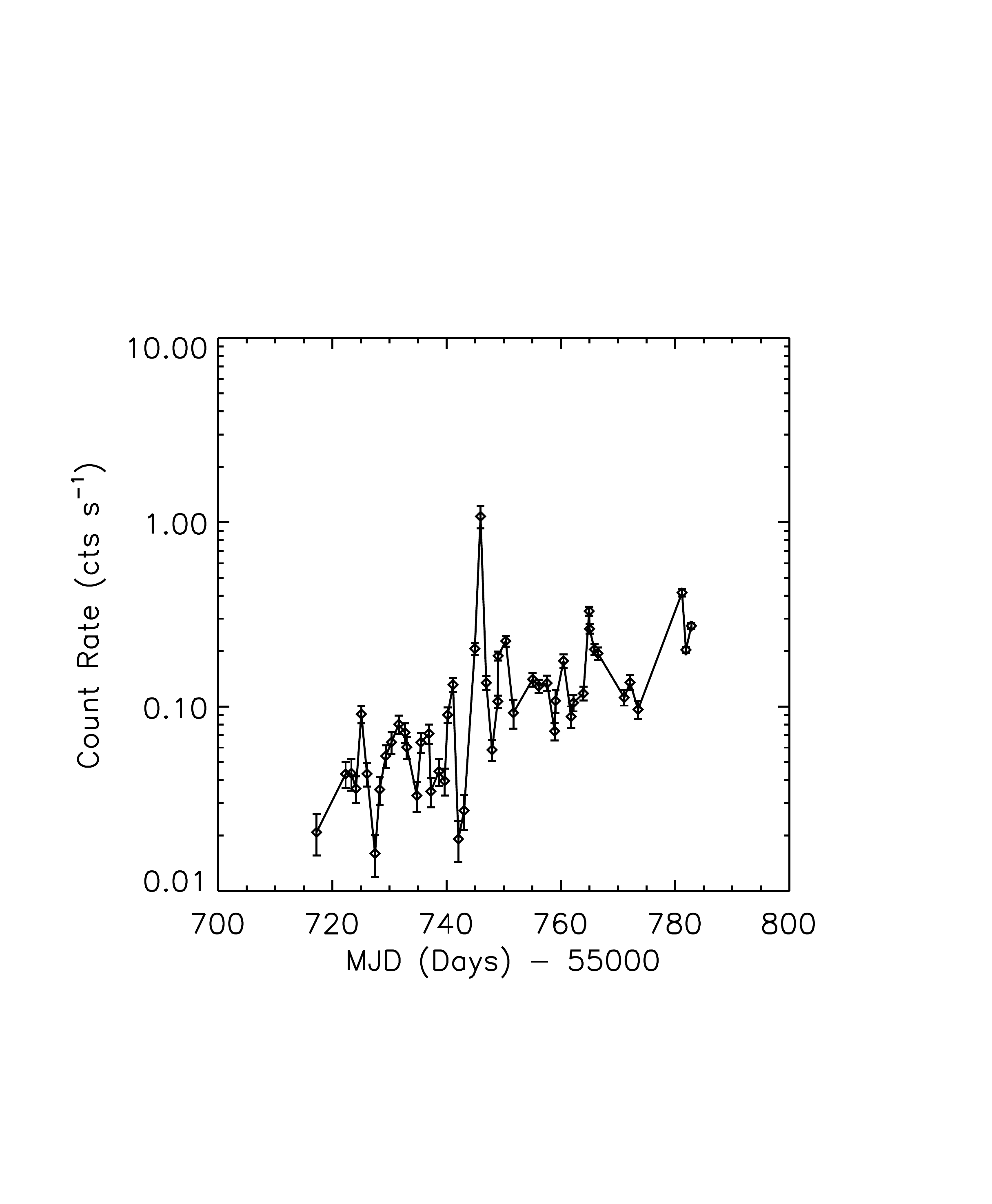}}
\subfigure[ \label{fig:colden}]{\includegraphics[scale=.44,angle=0,clip=true,trim= .5cm 5cm 5cm 3cm]{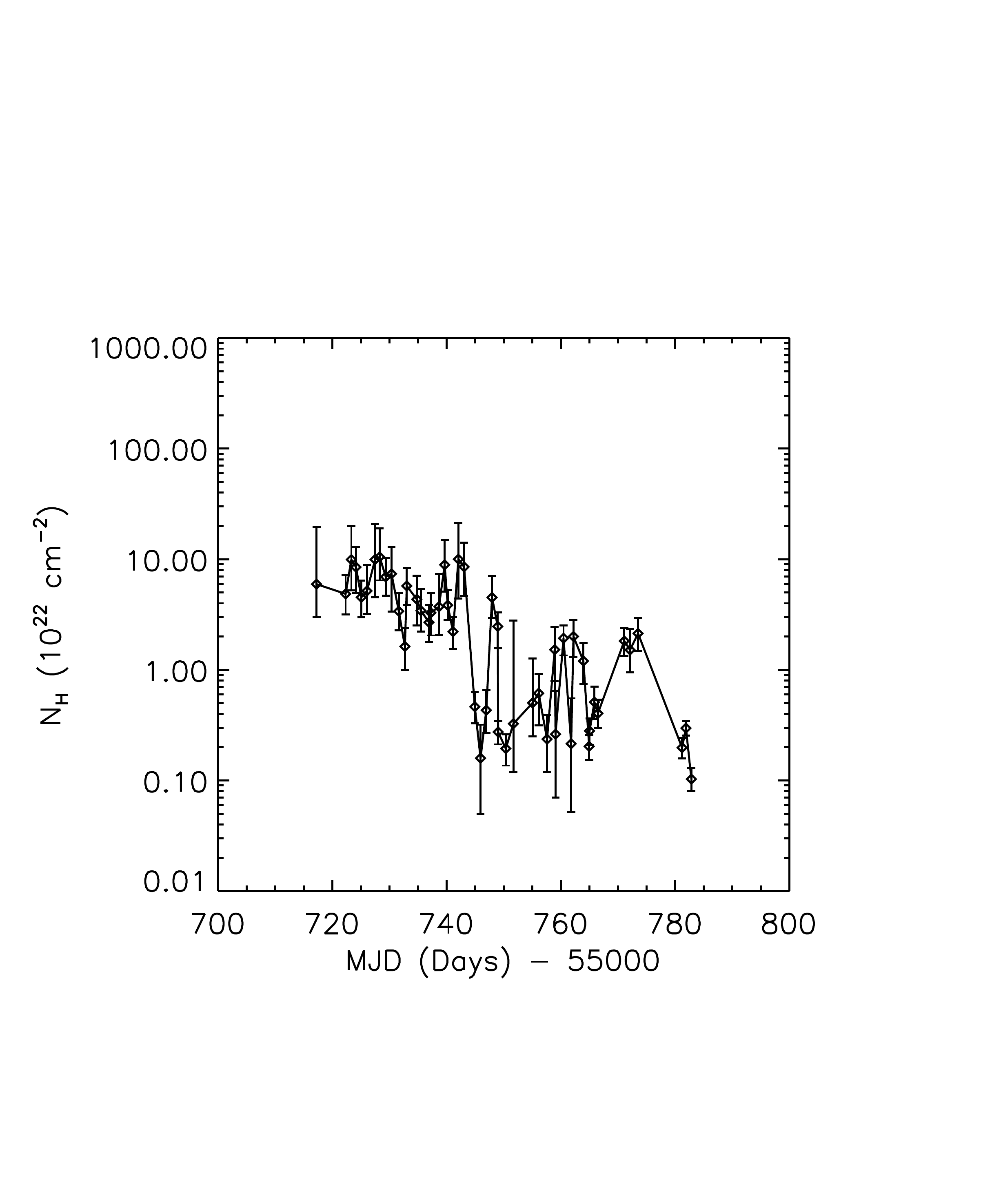}}
\subfigure[ \label{fig:nh}]{\includegraphics[scale=.4,angle=0,clip=true,trim= .5cm 4cm 5cm 0cm]{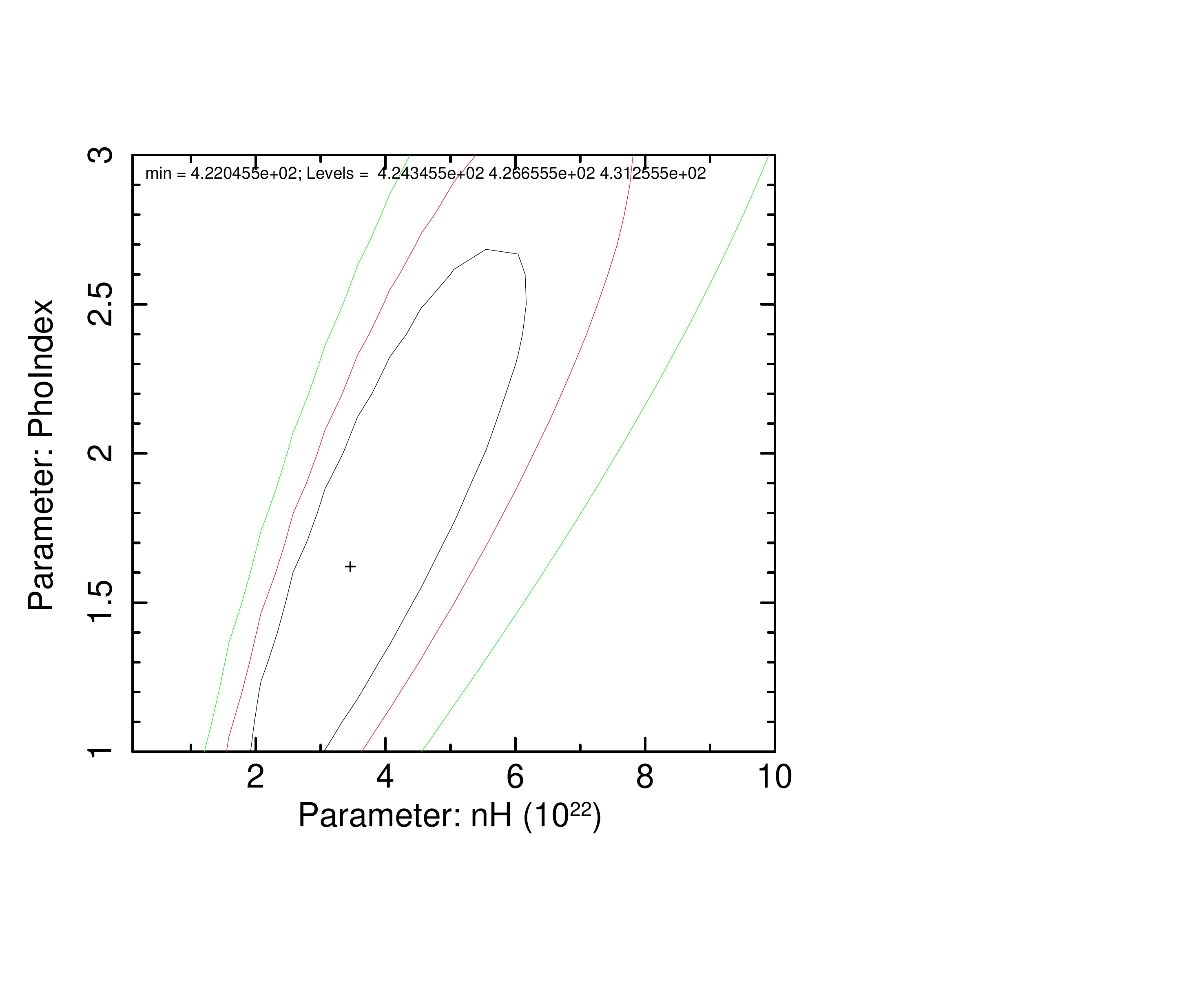}}
\caption{a) This plot shows the total (absorbed) count rate from the {\it Swift} observations. An increase in total count rate is observed throughout the campaign. This is consistent with the total column density changing by nearly two order of magnitude (See Figure \ref{fig:colden}), rather than the intrinsic continuum changing by the same order of magnitude (See Figure \ref{fig:lc}). b) This plot shows the column density (N$_H$) varying throughout the course of the observations. c) This plot shows the confidence contours when the column density and spectral index are allowed to vary in one particular observation (mjd=55740.2), with contours at 1, 2, \& 3 $\sigma$ confidence levels. The column density is constrained to be a relatively high column density.   }
\end{figure*}

The 50 X-ray observations were taken with {\it Swift}  in the photon counting mode on 5 June 2011 to 9 August 2011 with an average spacing of 1.3 days. The observations had approximately 1 ks exposures, and an average count rate of 0.13 cts s$^{-1}$, with a minimum count rate of 0.016 cts s$^{-1}$, and a maximum count rate of 1.08 cts s$^{-1}$ in the 0.3-10 keV band. See Figure \ref{fig:ctrate}.

To analyze the data, we used {\tt XSPEC} (v.\ 12.8.0) \citep{Arnaud99} and {\tt FTOOLS} (v.\ 6.13) \citep{Blackburn95}. We used the standard {\tt xrtpipeline} with the current response files from {\tt CALDB} (v. \ 4.5.1). Exposure maps were generated with {\tt xrtexpomap} and ancillary response matrices were made with {\tt xrtmkarf}. The ungrouped spectra were fit to a phenomenological power-law and two absorption components ({\tt tbabs$\ \times$ tbabs$\ \times$ power-law}) in {\tt XSPEC} using {\it cstat} statistics \citep{Cash79}. The average cstat per degree of freedom is $cstat/dof = 0.40$. The background was included as an annulus around the point source with inner radius of 140'' and outer radius of 210'', but in general the position and size of the background did not strongly affect our fits.  The first absorption component was frozen at N$_H$=1.85$\times 10^{20}\ \mathrm{cm}^{2}$ \citep{Kalberla05}, the Galactic column density, while the second absorption component was allowed to vary to model the neutral absorption within NGC 4395 (See Figure \ref{fig:colden}). Although the neutral absorption is highly variable, it primarily affects the low energy part of the spectrum, thus limiting the impact in the unabsorbed flux measurements in the 2--10 keV range we are concerned with. 

We  froze the power-law spectral index at $\Gamma =1.8$ due to the small number of counts, while still varying the normalization. Although some of the literature on NGC 4395 finds a much lower spectral index of $\Gamma \lesssim 1.5$ \citep[e.g.,][]{Lira99,Shih03,Moran05,Iwasawa10}, a higher spectral index, $\Gamma \approx 2.0 \pm0.2$, was found by \cite{Nardini11} when fitting an averaged {\it Swift/BAT} spectrum above 10 keV. This suggests that the hard spectral index at low X-ray energies is partly due to absorption and not intrinsic. Further, in Figure \ref{fig:nh} we show the confidence contours when we allow the spectral index and column density to both vary in one particular observation. The power-law index is consistent with a wide range of values, including $\Gamma = 1.8$ at better than 1$\sigma$ confidence, and the column density is constrained to a high column, i.e. $>1\times10^{22}cm^{-2}$. Finally, we note that when we fit our {\it Swift} data with $\Gamma = 1.5$, the resulting variability analysis does not change within 1 $\sigma$ statistical errors, as we are mainly concerned with the unabsorbed flux between 2--10 keV. 


Important to our campaign is the unabsorbed continuum flux which is shown in Figure \ref{fig:lc}, as the flux divided by the average flux. This shows the amount of intrinsic variability of the X-ray emission, which can be easily compared to the radio variability shown in Figure \ref{fig:rlc}.

\section{Analysis}
\subsection{Cross-Correlation Timing Analysis}
The variability in this data set is primarily seen in the X-ray and has a standard deviation, $\sigma$, divided by the mean, $\langle F_{X}\rangle$, of $\sigma/\langle F_X\rangle=0.51$ (Figure \ref{fig:lc}) The radio observations are less variable but do show statistically significant fluctuations. Three data points are above $2\sigma$ deviations from the mean, with the very last data point deviating from the mean by over 3.3$\sigma$. See Figure \ref{fig:rlc}. Using a normal distribution of fluctuations, one would expect only one data point to be above a $2\sigma$ deviation out of 16, suggesting the fluctuations we do see are real. In the radio data, the standard deviation divided by the mean flux-density is $\sigma/ \langle F_R\rangle=0.12$. These standard deviations are each dominated by the intrinsic variability rather than the measurement uncertainties.

\begin{figure*}[t]
\centering
\vspace{3cm}
\includegraphics[scale=.44,angle=0,clip=true,trim= 1cm 5cm 3cm 0cm]{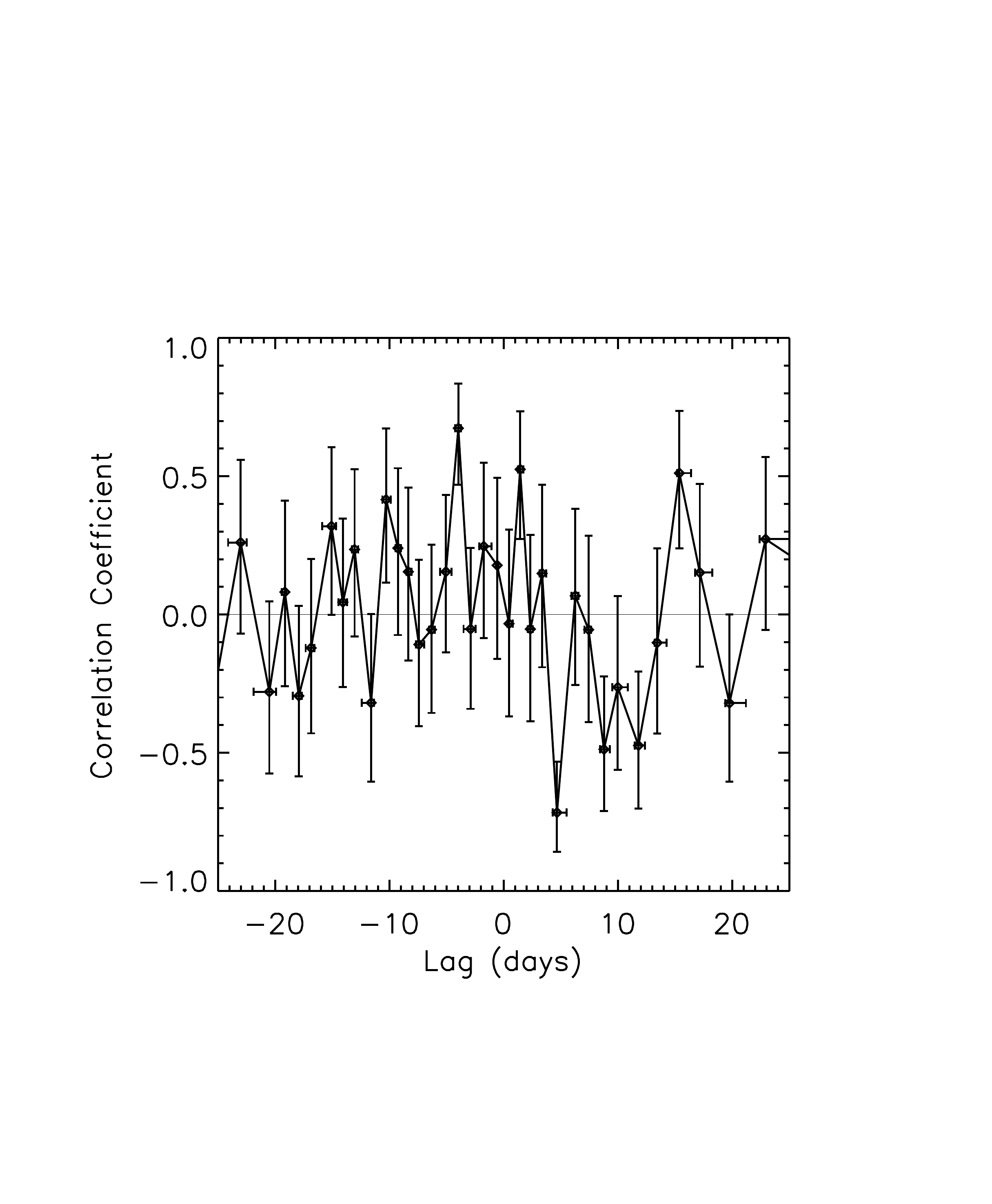}
\caption{This plot shows the z-transformed discrete correlation function for the X-ray fluxes versus radio flux-densities \citep{Alexander13}. We do not find any evidence of a statistically significant ($>5\sigma$) delay in the times series. \label{fig:dcf}}
\end{figure*}

We also checked for any apparent time-lags between the X-ray flux and the radio flux-density variability, which could influence our correlation analysis. We used the z-transformed discrete correlation function with a minimum of 11 data points per bin to determine a time lag \citep{Alexander13}. See Figure \ref{fig:dcf}. We find no evidence of a time lag, as there were no statistically significant correlation coefficients different from zero at a 5$\sigma$ confidence level. This was determined by dividing each correlation coefficient by its minus side error.


\subsection{X-ray vs. Radio Correlation}
Because the data are consistent with no time lag, we used the nearly simultaneous X-ray and radio observations in our correlation analysis. The average temporal separation between the X-ray and radio observations was 0.55 days, with a maximum separation of 1.6 days.  When correlating the data, we find a Spearman's $\rho$ correlation coefficient of $\rho=0.18$ with a null probability of $p=0.51$, and a Kendall's $\tau$ correlation coefficient of $\tau=0.15$ with a null probability $p=0.42$. The correlation test indicates that the weak correlation is not statistically significant. 

\begin{figure*}[t]
\centering
\vspace{1cm}
\subfigure[ \label{fig:FP}]{ \includegraphics[scale=.44,angle=0,clip=true,trim= 1cm 5cm 3cm 0cm]{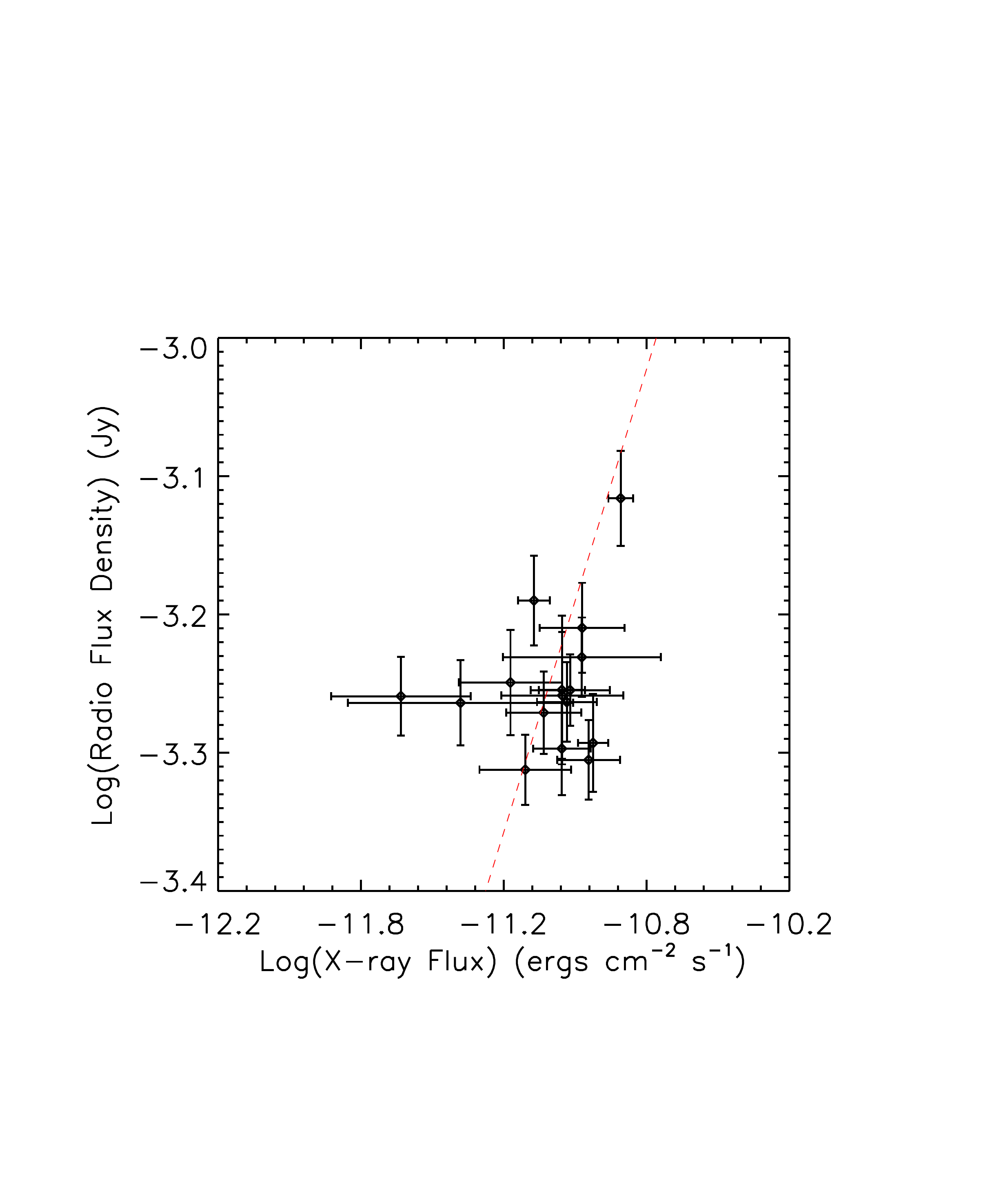}} 
\subfigure[ \label{fig:hist}]{\includegraphics[scale=.44,angle=0,clip=true,trim= 3cm 12cm 3cm 0cm ]{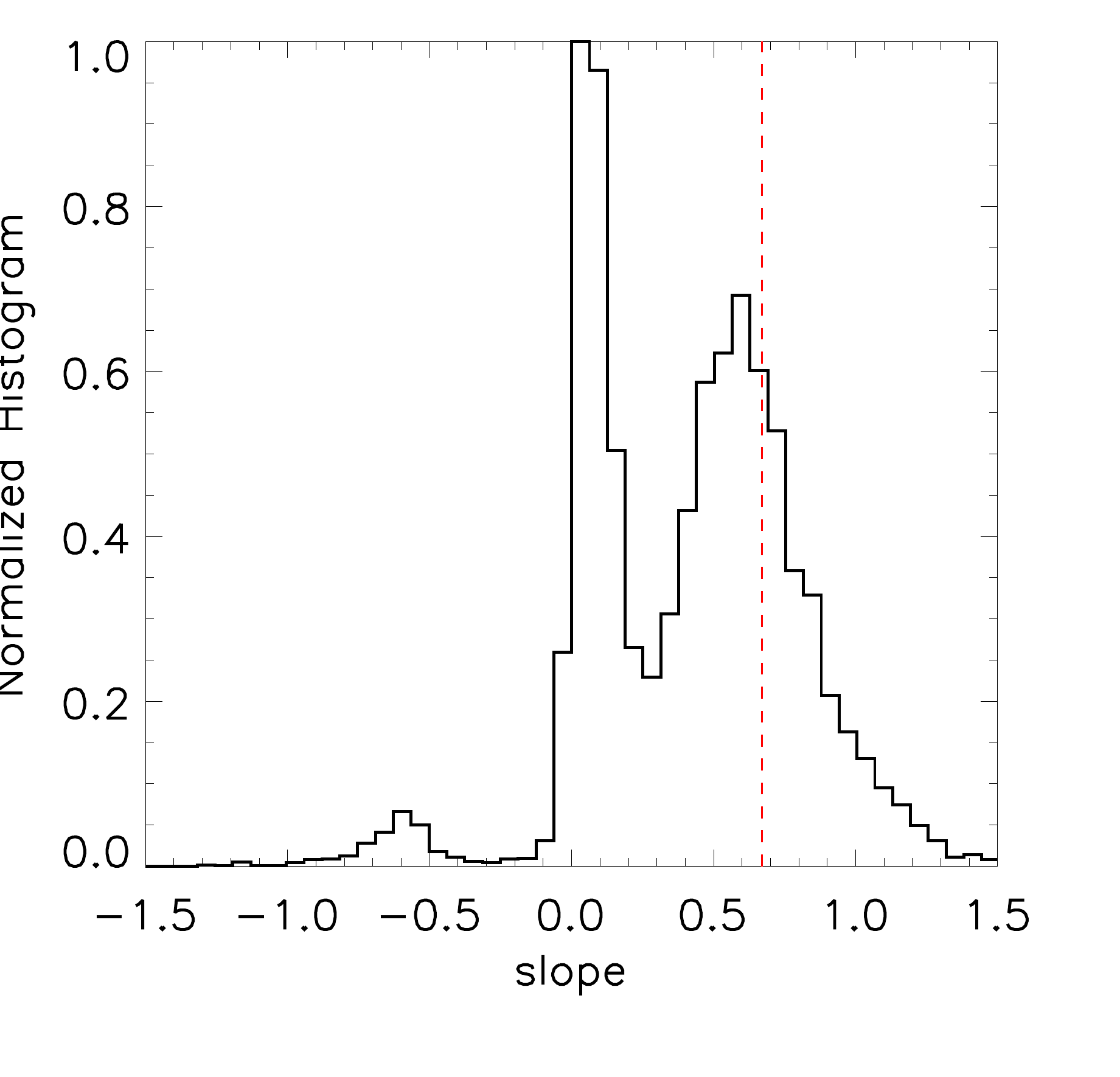}}
\caption{a) This plot shows the X-ray flux versus the radio flux-density. The red dashed line has a slope of the fundamental plane of black hole activity \citep{Gultekin09}. b) This is a histogram of the slopes from a bootstrap of N=10$^4$ resampling of the data shown in Figure \ref{fig:FP}. The peak is at $m=$0.06.  The slope is driven by the the lowest X-ray flux at $\log F_{X-ray} = - 11.61$ and the highest X-ray flux at $\log F_{X-ray} \approx - 10.84$ as shown in Figure \ref{fig:FP}. This is evidenced by the two main peaks at $m=0.06$ and $m=0.60$ and the small tail at $m<0$. The red dashed line is the slope 0.67 of the FP \citep{Gultekin09}. }
\end{figure*}

The X-ray flux versus the radio flux-density is plotted in Figure \ref{fig:FP}. As the ranking correlation suggested only a weak correlation, we fit the data with a linear relation. We found that the data are consistent with a flat relation but were also consistent with the fundamental plane given by the red dashed line. We used a bootstraping analysis to determine the linear-fit coefficients. Figure \ref{fig:hist} shows the normalized histogram of the 10$^4$ bootstrap re-samplings of the data, and the resulting slope, $m$, from these linear fits. There is a peak in the distribution at $m=0.06$, a second broad peak at $m=0.6$, and a small tail of the distribution at slopes less than 0. This shows that the data favor a flat slope, but are also consistent with the fundamental plane slope, $m=0.67$ \citep{Gultekin09}.

\section{Discussion}
\label{Disc}
The goal of this study was to examine the disk-jet coupling in an accreting SMBH. We chose to probe the viscous timescale of the inner accretion disk, as jets are thought to be launched within this region \citep[eg.,][]{Doeleman12}. In SMBHs, this timescale is on the order of a few days to months, making a simultaneous X-ray and radio observing campaigns feasible. Conversely, in stellar-mass black holes, this timescale is only a few 10's of seconds. 

In particular, we chose NGC 4395 because it is a bright, nearby Seyfert with the lowest known mass, making the appropriate cadence of such a campaign only a few days. We used {\it Swift} and the VLA to monitor both the X-ray from 2--10 keV and radio at 8.4 GHz over a two month period. 

In detail, we found that the X-ray variability was dominated by neutral absorption, but both the intrinsic X-ray continuum and compact radio emission did show variability. There was no statistically significant time delay between the X-ray and radio variability. We also correlated the nearly simultaneous X-ray and radio observations using a Spearman's ranking correlation test and found a weak positive correlation, $\tau=0.15$. Further, a linear fit to the data gave the relation $\log(L_R)=0.06 \log(L_X)+32.6$, but the data were also consistent with the fundamental plane of black hole activity with a slope of $m=0.67$. See Figures \ref{fig:FP} \& \ref{fig:hist}. This is consistent with the idea that the amplitude of the X-ray variability is greater than the radio variability, and the latter is responsible for driving NGC 4395 horizontally in the $L_X$-$L_R$ plane. 

NGC 4395 is not the only Seyfert that shows this general behavior. In our previous work, we show that NGC 4051 has higher variability in the X-ray as compared to its simultaneous radio observations \citep{King11}.  NGC 4051 is slightly larger at 1.73$^{+0.55}_{-0.52}\times10^6 M_\odot$ \citep{Denney10}, and the simultaneous X-ray and radio observing campaign of NGC 4051 probed the same viscous times as in NGC 4395 \citep{King11}. 
 
\begin{figure*}[t]
\centering
\vspace{3cm}
\subfigure[\label{fig:fp}]{\includegraphics[scale=.44,clip=true,trim= 1cm 10cm 5cm 2cm]{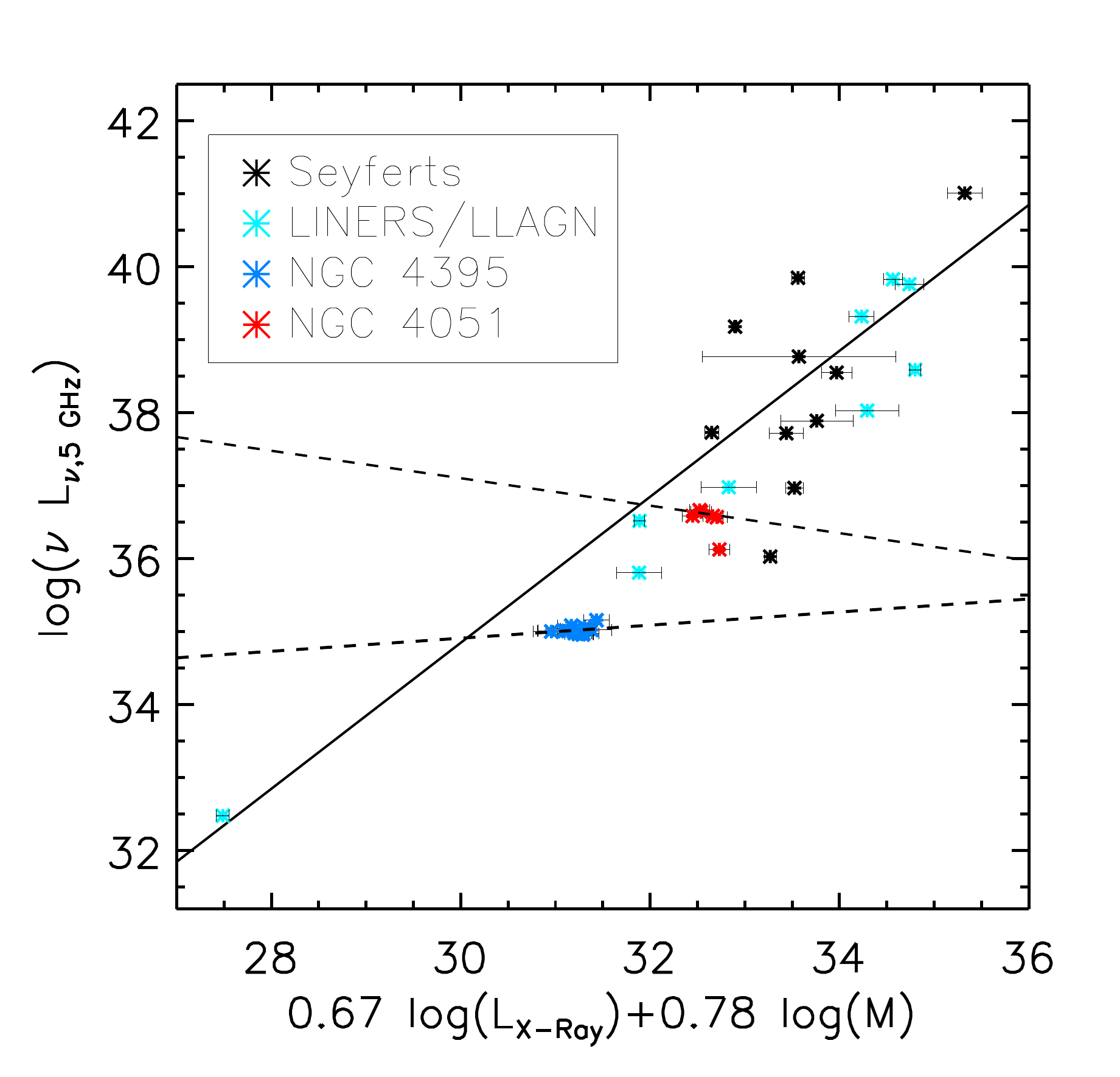}}
\subfigure[ \label{fig:fp_xrb}]{\includegraphics[scale=.44,angle=0,clip=true,trim= 1cm 5cm 5cm 5cm]{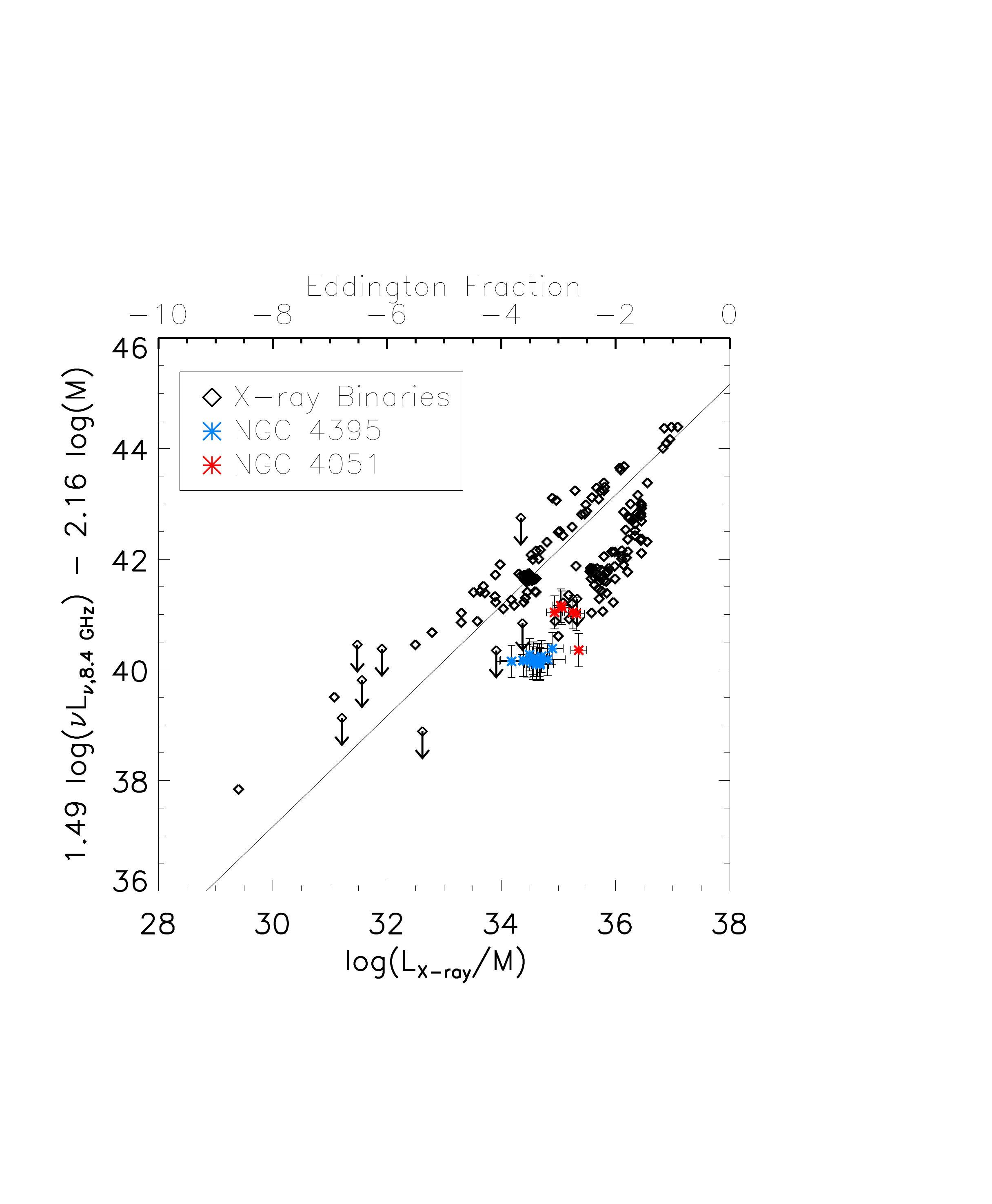}}
\caption{a) The above plot shows both NGC 4395 and NGC 4051 as they lie on the fundamental plane of black hole activity measured by \cite{Gultekin09}. The solid line in both Figure \ref{fig:fp} \& \ref{fig:fp_xrb} shows the plane derived by \cite{Gultekin09}, $\log(\nu L_{\nu,5GHz}) = 4.8 +0.78\log(M_{BH}) +0.67\log{L_X}$. The radio observations were converted to 5 GHz assuming $F_\nu \propto \nu^{-1}$. The dashed line shows the best fit lines to each of the Seyferts. In general NGC 4395 and NGC 4051 lie on the fundamental plane but move out of it when looking at simultaneous X-ray and radio observations on viscous timescales of the inner disk. b) This plot shows NGC 4395 and NGC 4051 plotted against the stellar-mass black holes as described in \cite{Gallo12}. The plot shows the Eddington ratio versus the radio luminosity corrected by the mass term, which is derived from the fundamental plane relation \citep[black line,][]{Gultekin09}. NGC 4395 and NGC 4051 appear to lie on the second, steeper track, which is suggestive that SMBH follow two distinct tracks just as stellar-mass black holes do.}
\end{figure*}

Shown in Figure \ref{fig:fp} are both NGC 4395 and NGC 4051, as they move out of the fundamental plane as given for Seyferts by \cite{Gultekin09}. Both sources are fairly constant in the radio, while the X-ray drives them out of the plane. Each of their respective best fit slopes are plotted as the dashed lines on Figure \ref{fig:fp}. As the two sources do lie on the plane, the X-ray variability may be partially responsible for the observed scatter of the fundamental plane.

In Figure \ref{fig:fp_xrb}, the two Seyferts are now plotted against the fundamental plane of stellar-mass black holes taken from work by \cite{Gallo12}. All the black holes have been corrected for mass using the fundamental plane derived from \cite{Gultekin09}, and a mass of 10$M_\odot$ has been assumed for the stellar-mass black holes. As noted in \cite{Gallo12}, the stellar-mass black holes occupy two different tracks: 1) the typical ``fundamental plane" track that scales a s $L_R\propto L_X^{0.63\pm0.03}$, and 2) a second track that is steeper that scales as $L_R\propto L_X^{0.98\pm0.08}$. NGC 4395 and NGC 4051 do lie on and may follow this steeper $L_X$-$L_R$ relation. If some SMBH also follow a second, steeper $L_X$-$L_R$ relation, it would imply that two distinct modes of accretion and jet production occur in both stellar-mass and SMBHs, and gives rise to a picture that the underlying physical mechanisms in disk-jet coupling scale across the black hole mass scale. 

Further, the flat slope in the $L_X$-$L_R$ plane traced by NGC 4395 and NGC 4051 suggests they are tracing a branch between the two tracks. This would be similar to the behavior of H1743-322, which jumps between the two tracks \citep{Coriat11,Gallo12}. However, the slope of NGC 4395 is also consistent with a steeper slope. In addition, as the amplitude of the X-ray variability is more variable than the radio variability, it is possible that strong variations in the disk or corona may get washed out when transferred to the jet on larger scales \citep[e.g.,][]{Maitra09}. In essence, the base of the jet might act as a low-pass filter for transferring only sufficiently large or sustained variations to the jet. This would imply that on short timescales and any Eddington ratio, a source would follow a flat relation and move out of the $L_X$-$L_R$ relation in Figure \ref{fig:fp_xrb}. On longer timescales variations the SMBH may trace out the fundamental plane, $L_R \approxprop L_X^{0.7}$ or the steeper relation of $L_R \approxprop L_X^1$. This is interesting because jets are known to be launched within 10's of gravitational radii \citep{Doeleman12}, which would correspond to viscous times of the inner disk. Yet our study points to longer timescales for disk-jet couplings, indicating global effects that propagate from further out in the accretion disk are vital to the disk-jet  coupling. 

Additional tracks in the $L_X$-$L_R$ plane have also been seen in not only stellar-mass black holes, but also in a few samples of low excitation galaxies (LEG) and Fanaroff-Riley I (FR I) galaxies \citep[e.g.,][]{Chiaberge02,Hardcastle06, Evans06, Hardcastle09, deGasperin11}. These sources generally fall above the conical ``fundamental plane", i.e. are radio bright or X-ray dim \citep{Hardcastle09,deGasperin11}. It has been suggested that the discrepancy between these sources is due to the LEG and FR I jet-dominated X-ray emission, while sources on the fundamental plane from \cite[e.g.,][]{Merloni03} have accretion dominated X-ray emission \citep{Chiaberge02,Evans06,deGasperin11}. Interestingly, NGC 4395, NGC 4051 and the second stellar-mass track do the opposite and lie below the ``fundamental plane", i.e. are X-ray bright and radio dim. This would argue that unlike the LEG and FR I sources whose X-ray emission is jet-dominated, that the X-ray emission is accretion-dominated like the fundamental plane sources but more efficiently radiating. This argument has also been suggested for AGN by \cite{Wu13}, who find low luminosity AGN form a second track as well, but in the $L_B$-$L_R$ plane where $L_B$ is the B band luminosity, another accretion disk proxy. 

\section{Conclusion}
In this study, we have observed NGC 4395 for approximately two months with nearly simultaneous 8.4 GHz radio observations with the VLA and 2--10keV {\it Swift} X-ray observations. We find that the X-ray flux has large variability that is dominated by variable neutral absorption. When we correlate the unabsorbed continuum X-ray variability with the radio variability, we find that the data are consistent with no time delay between the two bands and that the X-ray variability dominates over the radio variability. In addition, the data are consistent with the slope of the fundamental plane but also with a flat slope in the $L_X$-$L_R$ plane.

The average X-ray and radio luminosity of NGC 4395 as well as NGC 4051 are consistent with lying on the fundamental plane of black hole activity, and the X-ray variability driving some of the observed scatter. Furthermore, both Seyferts appear to lie on the ``second" $L_X$-$L_R$ stellar-mass black hole track discussed in \cite{Gallo12}. These sources may be probing a second, distinct disk-jet coupling, which is also seen in stellar-mass black holes. Our future work will be to probe higher accretion regimes in SMBHs to see if at higher Eddington ratios the SMBHs follow one of the two tracks or still move out of the plane like NGC 4395 and NGC 4051. In addition, longer timescales will be examined in order to assess global accretion affects.

\begin{acknowledgements}
We would like to thank the anonymous referee for their invaluable comments, as well as Michael Rupen for his instrumental help. A.L.K. gratefully acknowledges support through the NASA Earth and Space Sciences Fellowship. The National Radio Astronomy Observatory is a facility of the National Science Foundation operated under cooperative agreement by Associated Universities, Inc.
\end{acknowledgements}
\pagebreak

\end{document}